\documentclass[twocolumn,aps,pre]{revtex4}
\usepackage{amsmath,amssymb}
\usepackage{graphicx}
\usepackage{natbib}

\begin{document}

\title{High refractive index immersion liquid for super-resolution 3D imaging using sapphire-based aNAIL optics}

\author{Junaid M. Laskar$^1$, 
P. Shravan Kumar$^2$,
Stephan Herminghaus$^1$, 
Karen E. Daniels$^3$,
Matthias Schr\"oter$^{1,4}$}

\affiliation{$^1$Max Planck Institute for Dynamics and Self-Organization
(MPIDS), 37077 G\"ottingen, Germany \\
$^2$Dept. of Physics, IIT Guwahati, Guwahati, Assam, India \\
$^3$Dept. of Physics, NC State University, Raleigh, NC, USA \\
$^4$Institute for Multiscale Simulation, Friedrich-Alexander-Universit\"at
Erlangen-N\"urnberg, Erlangen, Germany}

\date{\today}

\begin{abstract}
Optically-transparent immersion liquids with refractive index ($n\sim 1.77$) to match sapphire-based aplanatic numerical aperture increasing lens (aNAIL) are necessary for achieving deep 3D imaging with high spatial resolution. We report that antimony tribromide (SbBr$_{3}$) salt dissolved in liquid diiodomethane (CH$_{2}$I$_{2}$) provides a new high refractive index immersion liquid for optics applications. The refractive index is tunable from $n=1.74$ (pure) to $n=1.873$ (saturated), by adjusting either salt concentration or temperature; this allows it to match (or even exceed) the refractive index of sapphire. Importantly, the solution gives excellent light transmittance in the ultraviolet to near-infrared range, an improvement over commercially-available immersion liquids. This refractive index matched immersion liquid formulation has enabled us to develop a sapphire-based aNAIL objective that has both high numerical aperture ($\mathrm{NA}=1.17$) and long working distance ($\mathrm{WD}=12$~mm). This opens up new possibilities for deep 3D imaging with high spatial resolution.
\end{abstract}

\maketitle

Copyright 2016 Optical Society of America. One print or electronic copy may be made for personal use only. Systematic reproduction and distribution, duplication of any material in this paper for a fee or for commercial purposes, or modifications of the content of this paper are prohibited.
The journal version can be accessed at 
https://www.osapublishing.org/ao/abstract.cfm?uri=ao-55-12-3165

\section{Introduction}

Sapphire-based aplanatic numerical aperture increasing lenses (aNAIL) \cite{ippolito2001high,ippolito2005theoretical} provide a promising route to achieve  super-resolution 3D imaging  \cite{serrels2008solid}, but their development requires a  suitable refractive index-matching liquid ($n \sim 1.77$).
A typical aNAIL design would include a truncated aplanatic solid immersion lens of plano-convex shape, made of high refractive index solid material \cite{mansfield1990solid,lu2013aberration,agarwal2015crossing}  such as sapphire \cite{wu1999imaging,lee2014vibrationally}. To date,  the lack of a suitable immersion liquid has limited the application of aNAIL to subsurface microscopy of objects immersed inside a refractive index-matched solid medium, without the possibility of depth-scanning \cite{ippolito2001high,ippolito2005theoretical,agarwal2015crossing}. 
Therefore, a refractive index-matched immersion liquid will allow for simultaneously harnessing both the high spatial 
resolution and the depth scanning capability of sapphire-based aNAILs \cite{serrels2008solid}. However, a persistent challenge in the search for high refractive index immersion liquids is to find one with both low absorbance and low scattering. The ideal liquid would provide optical transparency across the full spectrum from  ultraviolet to near-infrared, as well as tunability to provide precise index-matching  \cite{deetlefs2006neoteric}.

A promising candidate solvent is the organic liquid diiodomethane (CH$_2$I$_2$), which is one of the liquids with highest known refractive index values ($n =  1.74$). While other high refractive index liquids exist (phenyldi-iodoarsine (C$_6$H$_5$AsI$_2$) with $n=1.85$ and selenium monobromide (Se$_2$Br$_2$) with $n=2.1$ \cite{meyrowitz1955compilation}), diiodomethane has the key advantage of being commercially available. In addition, diiodomethane is an excellent solvent, and many liquid formulations using salts dissolved in 
diiodomethane are reported to increase the refractive index \cite{meyrowitz1955compilation} and are even available commercially (Cargille labs, Series M, $n = 1.8$). 
However, the strong light scattering and high absorbance of these formulations render them insufficiently transparent for high-resolution optics applications. 
A lack of knowledge of a salt formulation to increase the refractive index while maintaining optical transparency has caused diiodomethane to remain under-utilized as a preferred immersion solvent liquid, despite its inertness with many minerals (including sapphire) \cite{meyrowitz1955compilation}.

The refractive index of an optical medium is typically proportional to its mass density, as described by the Lorentz-Lorentz equation \cite{lamelas2013index}; this suggests that salts containing heavy elements would be promising candidates. In addition, a large electronegativity difference between the salt cation and anion typically predicts improved solubility. Guided by these principles, we screened  four different salts -- antimony tribromide  (SbBr$_{3}$) \cite{meyrowitz1955compilation}, antimony trichloride (SbCl$_{3}$), barium chloride (BaCl$_{2}$), and bismuth trichloride (BiCl$_{3}$) -- as potential solutes in CH$_2$I$_2$.
Of this list, BaCl$_{2}$ and BiCl$_{3}$  were found to have poor solubility due to small cation-anion electronegativity differences. 
Despite the excellent solubility of SbCl$_{3}$, the refractive index increased only by 0.02, due to the low atomic weight of chlorine. Therefore, only SbBr$_{3}$ was found to be a suitable candidate: for concentrations between 20 wt\% and saturation in liquid CH$_2$I$_2$, it achieves a refractive index as high as  $n\sim 1.873$. 
In addition, this solution shows excellent optical transparency and low scattering in the wavelength ranges $\lambda < 350~\mathrm{nm}$ and $450~\mathrm{nm} < \lambda < 1060~\mathrm{nm}$. Below, we present measurements quantifying the concentration, temperature, and wavelength dependence of the index of refraction, transmittance, and scattering of these liquid solutions.

\section{Methods}

We prepare liquid solutions of different concentrations (wt\%) from a powder sample of SbBr$_{3}$ (99\% pure, Alfa Aesar) dissolved in the liquid CH$_2$I$_2$ (99\% pure, Sigma-Aldrich). To obtain a solution of given wt\%, we mix the two components in the desired weight ratio, first on an electrical shaker for six hours at $T=22 \, ^\circ$C and then in a centrifuge for 3 hrs at 800 rpm at $T=15 \, ^\circ$C. We then separate the upper (supernatant) liquid solution from the precipitate (assumed to contain chemical impurities) collected at the bottom of the centrifuge tube. 
For a 50 wt\% concentration at room temperature ($T=22 \, ^\circ$C), crystals are also formed that remain in equilibrium with the supernatant (saturated) solution. Note that the chemical handling of both components requires significiant care. Antimony tribromide is harmful if swallowed (H302) or inhaled  (H332), and is is hygroscopic (absorbs water). Diiodomethane is harmful if swallowed (H302), causes skin 
irritation (H315), and serious eye damage (H318), and also may cause respiratory irritation (H335). 

\begin{figure}
\includegraphics[width=\linewidth]{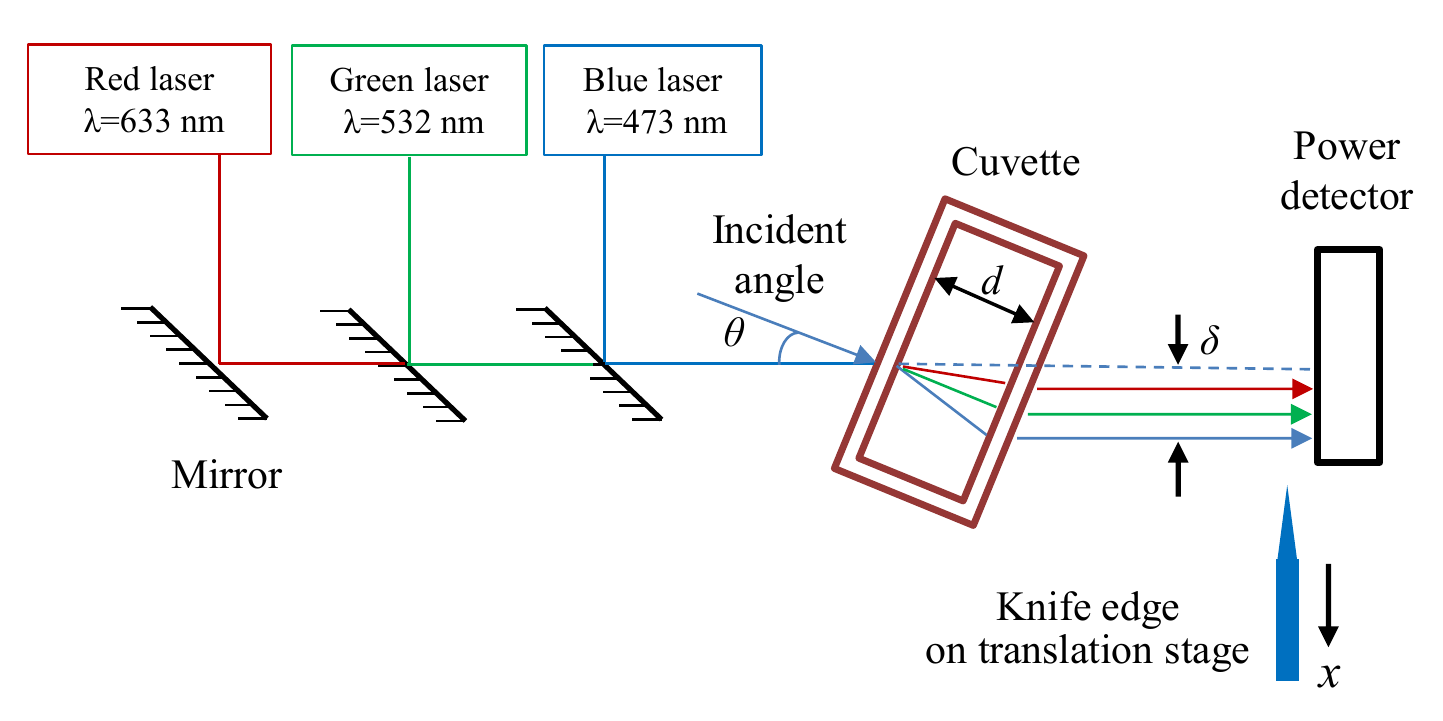}
\caption{Schematic of the liquid refractometer setup, based on the design of Nemoto
 \cite{nemoto1992measurement}.
The cuvette cross section is $20$~mm $\times
10$~mm, with optical measurements performed across the width $d=10$~mm.
}
\label{f:apparatus}
\end{figure}

\begin{figure}
\includegraphics[width=\linewidth]{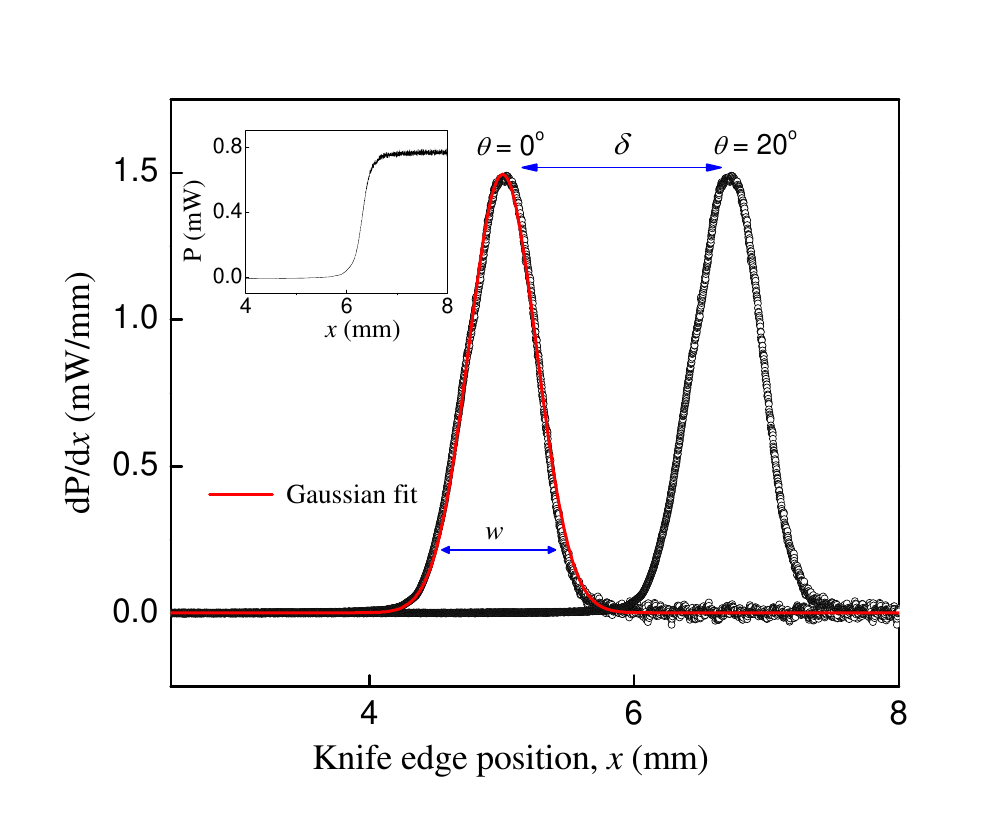}
\caption{Measurement of laser beam displacement  $\delta$  by knife-edge scanning. 
The inset shows the measured power versus the knife position. The main plot
shows fits to two Gaussian beam profiles (i.e.~derivatives of the inset)  for the angles
$\theta = 0^\circ $ and $20^\circ$. The difference between the two maxima is the displacement $\delta(20^\circ)$.}
\label{f:beam_profiling}
\end{figure}

We perform index of refraction measurements using a liquid refractometer based on the design of Nemoto \cite{nemoto1992measurement}, as shown schematically in in Fig.~\ref{f:apparatus}. The apparatus determines the displacement $\delta$ of a laser beam, due to passing through a liquid-filled cuvette rotated by an angle $\theta$ with respect to the beam. The center of the laser beam is determined by scanning a knife edge across its profile. As shown in the inset of Fig.~\ref{f:beam_profiling}, the light intensity profile $P(x)$ is well-described by an error function, as expected for a single-mode laser. We identify the center of the beam as the location $x$ at which $P(x)$ rises fastest, as obtained by numerical differentiation.  Sample Gaussian beam profiles $dP/dx$ are shown in Fig.~\ref{f:beam_profiling}, characterized by the beam width $w$ at which the power falls by a factor of $1/e^{2}$ \cite{chapple1994beam,magnes2006quantitative}. 
Throughout the measurements, we control the temperature of the liquid to $\pm 0.05^\circ$C by encapsulating the quartz cuvette (Hellma Analytics) inside a rectangular copper block containing channels connected to a refrigerated water bath circulator (Neslab RTE 111). The copper block is mounted on a rotation stage with a vernier scale which provides an angular reproducibility of $\pm 5$~arcmin.

Following Nemoto \cite{nemoto1992measurement}, the refractive index $n$ of the liquid can be calculated from
\begin{equation}
n = n_0 \sin{\theta} \sqrt{ 1 +
\left[ \frac{\cos{\theta}}{\sin{\theta}-\Delta/d} \right]^2 }
\label{eq:RI}
\end{equation}
where $n_0$ is the refractive index of air (the empty cuvette), $d$ is the width of the cuvette, and $\Delta \equiv \delta-\delta_0$ is the relative displacement of the Gaussian peak for the liquid-filled cuvette relative to the empty cuvette.

To obtain $n$, we repeat the same measurement at 7 different incident angles
$\theta = (\pm10^\circ, \pm20^\circ, \pm30^\circ, +40^\circ)$, corresponding to both  clockwise and counter-clockwise rotation of the cuvette with respect to the incident light. The average of these 7 measurements provides the value of $n$ for each combination of wavelength, temperature, and SbBr$_3$ concentration. We repeat this process for 3 wavelengths of light ($\lambda = 473, 532,$ and $633$~nm), 5 temperatures from $15^\circ$C to $40^\circ$C, and 3 concentrations (20 wt\%,  33.5 wt\%, and saturation concentration).

Eq.~\ref{eq:RI} already reduces systematic errors due to geometric imperfections of the cuvette by measuring all values of $\delta$ against the empty cuvette. In addition, we have analyzed the  systematic errors in our setup and determined that the precision of the rotation stage and the imperfect parallelism of the cuvette sidewalls are the largest sources of error. These combined effects result in a refractive index measurement error of $\pm 0.003$. In practice, we find that we are able to measure the refractive index of water and ethanol to within $\pm0.001$ of values reported in the literature  \cite{schiebener1990refractive, zaidi1989accurate, moreels1984laser}.

\section{Results}
\subsection{Index of Refraction}

\begin{figure*}
\includegraphics[width=18.8 cm]{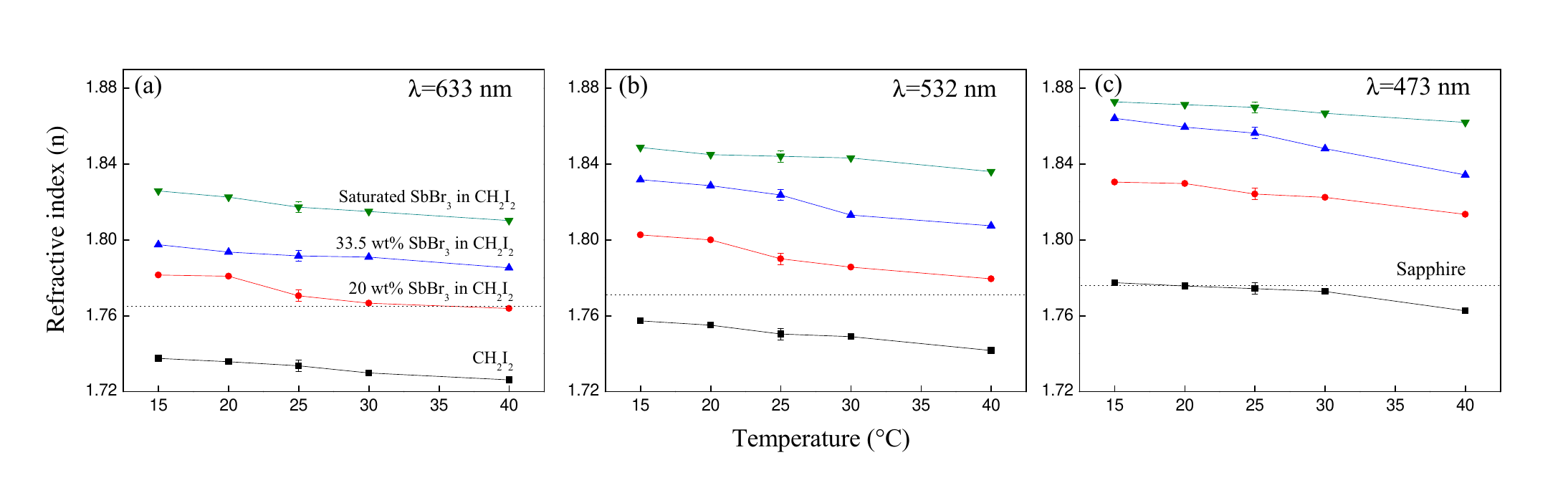}
\caption{Refractive indices of both pure diiodomethane (black lines) and with dissolved antimony tribromide (colored lines). Each panel compares the results for a single wavelength, as a function of concentration and temperature. The refractive index for ordinary rays through sapphire is shown by dotted line for comparison.
}
\label{f:RI_vs_Temp}
\end{figure*}

We find that SbBr$_{3}$ dissolved in CH$_2$I$_2$ can meet and exceed the index of refraction of sapphire over a broad range of temperatures, visible wavelengths, and concentration greater than 20 wt\%. Fig.~\ref{f:RI_vs_Temp} presents the measured values of $n$ as a function of wavelength (panels a-c), temperature (plot abscissa), and concentration (line series). In each case, the refractive index can be increased by either changing the concentration (more dissolved salt corresponds to higher $n$) or the temperature (higher temperature decreases $n$). In applications, preparing a solution of known concentration is more convenient for coarse tuning, and temperature is more convenient for fine tuning in-situ. The largest value we measure is $n=1.873$, for saturated solutions at low temperature and short wavelength.

\subsection{Optical Transparency}

\begin{figure}
\includegraphics[width=\linewidth]{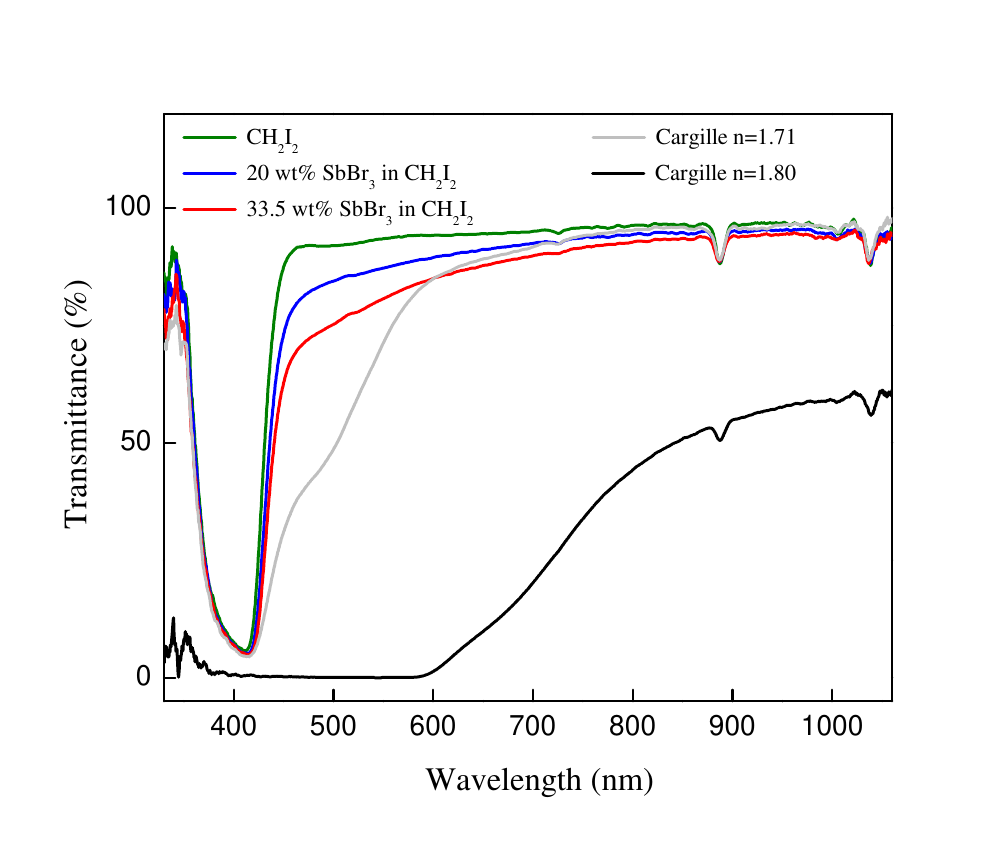}
\caption{Transmittance  as a function of wavelength for a cuvette of width $d=10$~mm.}
\label{f:Transmittance_spectra}
\end{figure}

Optical transparency, which can be degraded by both light scattering and absorption, plays a crucial role in determining the utility of an immersion liquid. We characterize these properties of the SbBr$_{3}$-CH$_2$I$_2$ solutions, and compare with the two commonly-used commercial high refractive index immersion liquids (Cargille M Series with $n=1.71 \pm 0.0005$ and $n=1.80 \pm 0.0005$). 
Fig.~\ref{f:Transmittance_spectra} shows the transmittance spectra, measured from the near-ultraviolet to the near-infrared. The illumination source is an Ocean Optics tungsten-halogen light source (model HL-2000-FHSA-LL with output power 4.5 mW) and transmitted light is recorded on an  Ocean Optics spectrometer (model HR2000+). For all five liquid solutions, there is a significant absorption band centered around $410$~nm (blue). In addition, there are several coincident absorption bands located at $\lambda=725, 887,$ and $1037$~nm, which possibly arise due to the common solvent CH$_2$I$_2$ used for all five liquids. 

Furthermore, we observe a Tyndall effect at all the three wavelengths, with stronger scattering for the Cargille liquids than for our SbBr$_{3}$ solutions. This suggests that  Mie scattering is present, caused by colloidal particles with a size on the same order as $\lambda$ \cite{bohren2008absorption}. 
In order to quantify the amount of scattering, we measure the relative increase in beam width $\Delta w = \frac{w- w_0}{w_0}$, where $w$ and $w_0$ are the beam-widths for the liquid and air-filled cuvettes, respectively. 
A large value of $\Delta w$ signifies stronger scattering. For simplicity, all measurements are done at $T=25^\circ$C and for normal incidence ($\theta =0^\circ$), using the knife-edge scanning method shown in Fig.~\ref{f:beam_profiling}. 
Results for the two wavelengths $473$ and $633$~nm are given in Table~\ref{t:scattering}.

\begin{table}
\begin{center}
\begin{tabular}{l|c|c}
Liquid/liquid solution & \multicolumn{2}{l|}{Beam width increase(\%)}  \\
\cline{2-3} &     $\lambda=633$~nm      &      $\lambda=473$~nm
\\ \hline
Diiodomethane     (CH$_2$I$_2$)  &  0  &   2.2 \\ \hline
20 wt\%  SbBr$_{3}$  solution   &    0 & 9.5 \\ \hline
33.5 wt\%  SbBr$_{3}$  solution   &  0 &   10.6  \\ \hline
Cargille liquid   ($n$=1.71)  &  1.2 &   16  \\ \hline
Cargille liquid   ($n$=1.80)  &  13.5 &   Opaque \\ 
\end{tabular}
\end{center}
\caption{Proportional increase (\%) in the beam width, referenced against an emtpy cuvette.}
\label{t:scattering}
\end{table}

At $\lambda=633~\mathrm{nm}$, the beam width remains unaffected for the SbBr$_{3}$-CH$_2$I$_2$ liquid solutions and the Cargille liquid $n=1.71$. 
Only the Cargille $n=1.80$ liquid contains colloidal particles in the relevant size range. At $\lambda=473$~nm, the beam width also shows a concentration-dependent increase for the SbBr$_{3}$-CH$_2$I$_2$ liquid solutions. 
A likely source of particles in this diameter range is that hydrolysis with atomspheric humidity produces small antimony oxide crystals 
via the reaction $2\mathrm{SbBr}_3 + 3 \mathrm{H}_2 \mathrm{O} \rightarrow \mathrm{Sb}_2 \mathrm{O}_3 + 6\mathrm{HBr}$ \cite{lide2012crc}.
At the same wavelength, the Cargille liquid $n=1.71$ shows an even stronger scattering, while the opacity of the $n=1.8$ Cargille liquid does not even allow the measurement of the beam width.

\subsection{Physical Properties}

We characterize the volumetric thermal expansion coefficient ($\gamma$)
and the density ($\rho$) of the liquid solution (SbBr$_{3}$+CH$_2$I$_2$) as a function of concentration, using a dilatometer (PHYWE Systeme GmbH) and pycnometer (BRAND GmbH),
respectively. We observe that decreasing the concentration of SbBr$_3$
from the saturated case increases $\gamma$ by  50\% and decreases $\rho$ by 4\%,
as shown in Table~\ref{t:Physicsal_properties}. Because SbBr$_3$ reacts with H$_2$O, the
liquid solution (SbBr$_{3}$+CH$_2$I$_2$) is immiscible in water (see Sec 3.B for details).

\begin{table}
\begin{center}
\begin{tabular}{l|c|c}
Liquid solution     &   \begin{tabular}{@{}c@{}}Thermal expansion \\ coefficient, $\gamma$ $(K^{-1})$\end{tabular}  &  \begin{tabular}{@{}c@{}} Density, $\rho$ \\ (gm/mL)\end{tabular} \\ \hline
20 wt\%  SbBr$_{3}$  solution   &   6.9$\times 10^{-4}$ & 3.422 \\ \hline
33.5 wt\%  SbBr$_{3}$  solution   &  5.4$\times 10^{-4}$ & 3.504\\ \hline
Saturated SbBr$_{3}$  solution &  4.6$\times 10^{-4}$ & 3.572  \\ 
\end{tabular}
\end{center}
\caption{Physicsal properties of the liquid solutions of different concentrations. All density measurements measurements were performed at $T=22 \, ^\circ$C. The temperature ranges, over which the volumetric thermal expansion measurements were performed, were $T=23-35 \, ^\circ$C, $20-32 \, ^\circ$C and $25-36 \, ^\circ$C for 20 wt\%, 33.5 wt\% and saturated solution, respectively.}
\label{t:Physicsal_properties}
\end{table}

\section{Discussion}

\begin{figure}
\includegraphics[width=\linewidth]{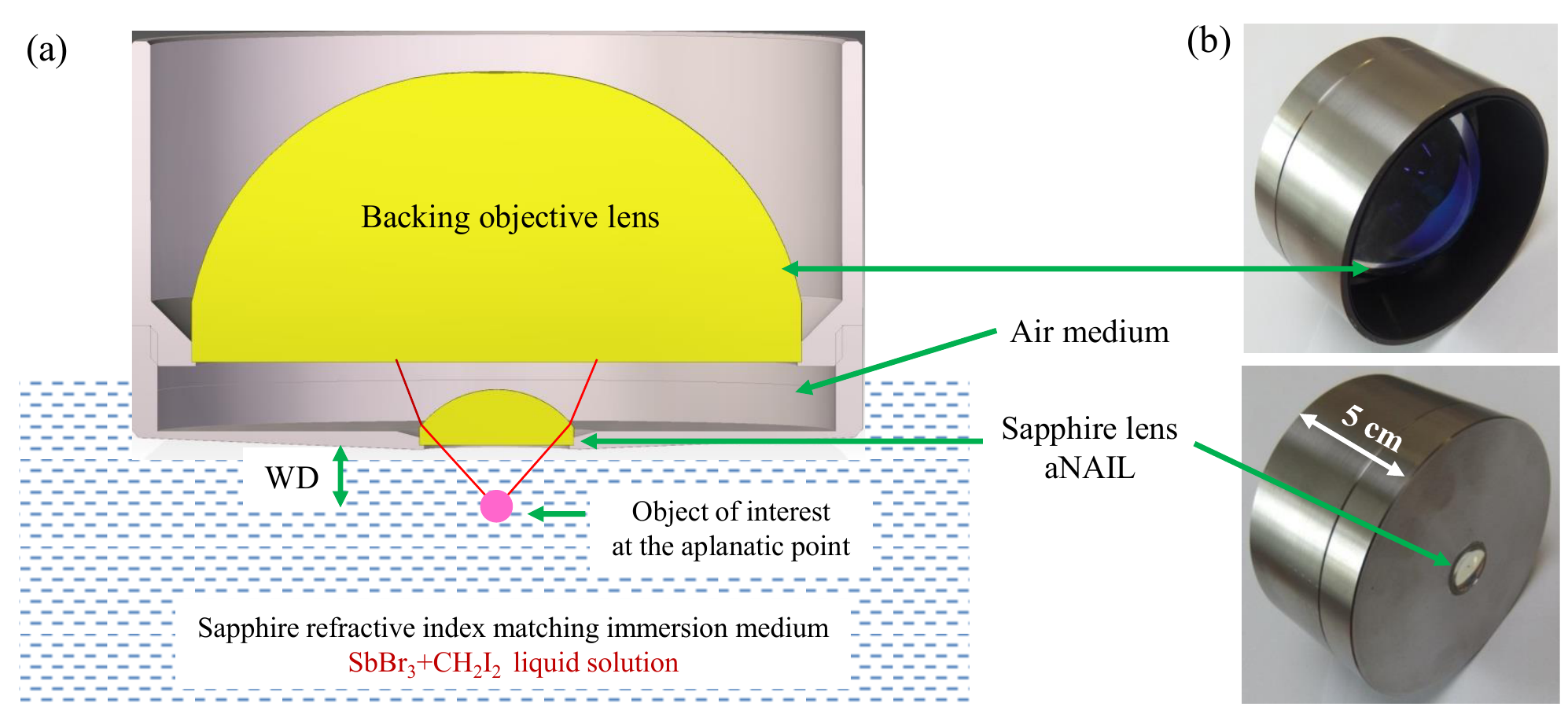}
\caption{(a) Schematic of the design for simultaneous increase of $\mathrm{NA}$ and $\mathrm{WD}$, using sapphire-based aNAIL lens system immersed in the refractive index matching liquid (solution of SbBr$_{3}$ in CH$_2$I$_2$). 
(b) Picture of an aNAIL and backing objective lens system, designed to have $\mathrm{NA}=1.17$ and $\mathrm{WD}=12$~mm.}
\label{f:lens}
\end{figure}

Antimony tribromide (SbBr$_{3}$) dissolved in diiodomethane (CH$_2$I$_2$) is a strong candidate as an immersion liquid for sapphire-based aNAIL lenses (see Fig.~\ref{f:lens}a). Together with a refractive index matched immersion liquid, these lenses allow for the simultaneous increase of both the numerical aperture ($\mathrm{NA}$, the light-gathering power) and the working distance ($\mathrm{WD}$) of an objective lens \cite{ippolito2005theoretical,serrels2008solid}. This technique places the object of interest at the aplanatic point of the spherical surface of the aNAIL lens in order to have an aberration-free focal spot. It can increase the $\mathrm{NA}$ of the backing objective lens by a factor of $n_{\mathrm{aNAIL}}^2$, up to the maximum achievable value $n_{\mathrm{aNAIL}}$\cite{ippolito2001high,ippolito2005theoretical}.

Fig.~\ref{f:lens}b shows the picture of the aNAIL and the backing objective lens system designed for significantly increasing the numerical aperture ($\mathrm{NA}$) of the backing objective lens from $0.619$ to $1.17$, while still maintaining a long working distance ($\mathrm{WD}=12$~mm). The lens system was designed using WinLens3D optical design software. 

Because $n$ is tunable via both concentration and temperature, the liquid formulation presented here could open new routes to creating adaptive lenses with tunable optical power \cite{ren2006tunable}. In these applications, liquid lenses have the advantage of additionally allowing the {\it shape} of the lens to be tuned in order to adjust its focal length (and therefore the optical power), mimicking the mechanism of human eye. To date, the lack of an appropriate high-$n$ liquid has limited the range of tunable optical power, as this depends on the difference of the refractive indices of the two immiscible liquids used to build the adaptive liquid lens \cite{li2011liquid, kuiper2004variable, mishra2014optofluidic}. 

Finally, we close with the application that inspired our work on this immersion liquid formulation: the imaging of gems made from corundum minerals (ruby and sapphire), which have $n \sim 1.77$. High spatial resolution lenses such as those described above (aNAIL) would open new routes to studying the crystallization process during mineral formation \cite{wenk2004minerals} and quality in manufacturing industry (gemstone, watch etc.) \cite{hughes1997ruby}. Importantly, ruby also has a pressure-dependent fluorescence peak  \cite{mao1986calibration, holzapfel2003refinement,syassen2008ruby}, which has recently been exploited as a tool for measuring the pressure field inside rubies themselves  \cite{chen2007stress,chen2006pressure}.

\section{Conclusion}

We have demonstrated that antimony tribromide (SbBr$_{3}$) dissolved in diiodomethane (CH$_2$I$_2$), for concentrations between 20 wt\% and saturation, provides a promising new high refractive index liquid formulation. At standard conditions for temperature and pressure (STP) and at wavelengths from near-ultraviolet to near-infrared, we observe that $n>1.77$ (sapphire) is attainable. For optimized choice of parameters (low temperature, short wavelength, high concentration), we can achieve $n=1.873$. In addition, this liquid is observed to have high transmission and low scattering for $\lambda < 350$~nm and $\lambda \gtrsim 450$~nm. On using this liquid formulation as refractive index matched immersion medium, we have developed a sapphire-based aplanatic numerical aperture increasing lens (aNAIL) that has both high numerical aperture $\mathrm{NA}=1.17$ and long working distance ($\mathrm{WD}=12$~mm). This paves the way  for deep 3D imaging with high spatial resolution. Moreover, the tunability of this extremely high $n$ liquid has several other promising applications, particularly in the development of adapative liquid lens, gemstone, watch industry etc. 

\paragraph*{Acknowledgments:} The authors would like to thank Markus Benderoth and Kristian Hantke for invaluable technical expertise. In addition, K.E.D. is grateful to the Alexander von Humboldt Foundation and the James S. McDonnell Foundation for their financial support. P.S.K. is grateful to the FOKOS society for providing him with a Ludwig Prandtl stipend.


\begin{thebibliography}{10}
\newcommand{\enquote}[1]{``#1''}

\bibitem{ippolito2001high}
S.~B. Ippolito, B.~Goldberg, and M.~{\"U}nl{\"u}, \enquote{High spatial  resolution subsurface microscopy,} Appl. Phys. Lett. \textbf{78}, 4071--4073  (2001).

\bibitem{ippolito2005theoretical}
S.~Ippolito, B.~Goldberg, and M.~{\"U}nl{\"u}, \enquote{Theoretical analysis of
  numerical aperture increasing lens microscopy,} J. Appl. Phys. \textbf{97},
  053105 (2005).

\bibitem{serrels2008solid}
K.~A. Serrels, E.~Ramsay, P.~A. Dalgarno, B.~Gerardot, J.~O'Connor, R.~H.
  Hadfield, R.~Warburton, and D.~Reid, \enquote{Solid immersion lens
  applications for nanophotonic devices,} J. Nanophotonics \textbf{2},
  021854--021854 (2008).

\bibitem{mansfield1990solid}
S.~M. Mansfield and G.~Kino, \enquote{Solid immersion microscope,} Appl. Phys.
  Lett. \textbf{57}, 2615--2616 (1990).

\bibitem{lu2013aberration}
Y.~Lu, T.~Bifano, S.~{\"U}nl{\"u}, and B.~Goldberg, \enquote{Aberration
  compensation in aplanatic solid immersion lens microscopy,} Opt. Express
  \textbf{21}, 28189--28197 (2013).

\bibitem{agarwal2015crossing}
K.~Agarwal, R.~Chen, L.~S. Koh, C.~J. Sheppard, and X.~Chen, \enquote{Crossing
  the resolution limit in near-infrared imaging of silicon chips: Targeting
  10-nm node technology,} Phys. Rev. X \textbf{5}, 021014 (2015).

\bibitem{wu1999imaging}
Q.~Wu, R.~D. Grober, D.~Gammon, and D.~Katzer, \enquote{Imaging spectroscopy of
  two-dimensional excitons in a narrow gaas/algaas quantum well,} Phys. Rev.
  Lett. \textbf{83}, 2652 (1999).

\bibitem{lee2014vibrationally}
E.~S. Lee, S.-W. Lee, J.~Hsu, and E.~O. Potma, \enquote{Vibrationally resonant
  sum-frequency generation microscopy with a solid immersion lens,} Biomed.
  Opt. Express \textbf{5}, 2125--2134 (2014).

\bibitem{deetlefs2006neoteric}
M.~Deetlefs, K.~R. Seddon, and M.~Shara, \enquote{Neoteric optical media for
  refractive index determination of gems and minerals,} New J. Chem.
  \textbf{30}, 317--326 (2006).

\bibitem{meyrowitz1955compilation}
R.~Meyrowitz, \enquote{A compilation and classification of immersion media of
  high index of refraction,} Am. Mineral. \textbf{40}, 398--409 (1955).

\bibitem{lamelas2013index}
F.~Lamelas, \enquote{Index of refraction, density, and solubility of ammonium
  iodide solutions at high pressure,} The Journal of Physical Chemistry B
  \textbf{117}, 2789--2795 (2013).

\bibitem{nemoto1992measurement}
S.~Nemoto, \enquote{Measurement of the refractive index of liquid using laser
  beam displacement,} Appl. Opt. \textbf{31}, 6690--6694 (1992).

\bibitem{chapple1994beam}
P.~B. Chapple, \enquote{Beam waist and m2 measurement using a finite slit,}
  Opt. Eng. \textbf{33}, 2461--2466 (1994).

\bibitem{magnes2006quantitative}
J.~Magnes, D.~Odera, J.~Hartke, M.~Fountain, L.~Florence, and V.~Davis,
  \enquote{Quantitative and qualitative study of gaussian beam visualization
  techniques,} arXiv preprint physics/0605102  (2006).

\bibitem{schiebener1990refractive}
P.~Schiebener, J.~Straub, J.~L. Sengers, and J.~Gallagher, \enquote{Refractive
  index of water and steam as function of wavelength, temperature and density,}
  J. Phys. Chem. Ref. Data \textbf{19}, 677--717 (1990).

\bibitem{zaidi1989accurate}
A.~Zaidi, Y.~Makdisi, K.~Bhatia, and I.~Abutahun, \enquote{Accurate method for
  the determination of the refractive index of liquids using a laser,} Rev.
  Sci. Instrum. \textbf{60}, 803--805 (1989).

\bibitem{moreels1984laser}
E.~Moreels, C.~De~Greef, and R.~Finsy, \enquote{Laser light refractometer,}
  Appl. Opt. \textbf{23}, 3010--3013 (1984).

\bibitem{bohren2008absorption}
C.~F.~Bohren, and D.~R.~Huffman, Absorption and scattering of light by small particles, \emph{ (John Wiley \&
  Sons, 2008)}.

\bibitem{lide2012crc}
D.~R.~Lide, Handbook of Chemistry and Physics: A Ready-reference Book of Chemical and
  Physical Data 2012-2013, \emph{ (CRC, 2012)}.

\bibitem{ren2006tunable}
H.~Ren, D.~Fox, P.~A. Anderson, B.~Wu, and S.-T. Wu, \enquote{Tunable-focus
  liquid lens controlled using a servo motor,} Opt. Express \textbf{14},
  8031--8036 (2006).

\bibitem{li2011liquid}
L.~Li, Q.-H. Wang, and W.~Jiang, \enquote{Liquid lens with double tunable
  surfaces for large power tunability and improved optical performance,} J.
  Opt. \textbf{13}, 115503 (2011).

\bibitem{kuiper2004variable}
S.~Kuiper and B.~Hendriks, \enquote{Variable-focus liquid lens for miniature
  cameras,} Appl. Phys. Lett. \textbf{85}, 1128--1130 (2004).

\bibitem{mishra2014optofluidic}
K.~Mishra, C.~Murade, B.~Carreel, I.~Roghair, J.~M. Oh, G.~Manukyan, D.~van~den
  Ende, and F.~Mugele, \enquote{Optofluidic lens with tunable focal length and
  asphericity,} Sci. Rep. \textbf{4} (2014).

\bibitem{wenk2004minerals}
H.~R.~Wenk and A.~Bulakh, Minerals: their constitution and origin, \emph{ (Cambridge University Press,
  2004)}.

\bibitem{hughes1997ruby}
R.~W.~Hughes, Ruby \& sapphire, \emph{ (Rwh Pub, 1997)}.

\bibitem{mao1986calibration}
H.~Mao, J.~Xu, and P.~Bell, \enquote{Calibration of the ruby pressure gauge to
  800 kbar under quasi-hydrostatic conditions,} J. Geophys. Res. \textbf{91},
  4673--4676 (1986).

\bibitem{holzapfel2003refinement}
W.~B. Holzapfel, \enquote{Refinement of the ruby luminescence pressure scale,}
  J. Appl. Phys. \textbf{93}, 1813--1818 (2003).

\bibitem{syassen2008ruby}
K.~Syassen, \enquote{Ruby under pressure,} High Pressure Res. \textbf{28},
  75--126 (2008).

\bibitem{chen2007stress}
Y.~Chen, A.~Best, T.~Haschke, W.~Wiechert, and H.-J. Butt, \enquote{Stress and
  failure at mechanical contacts of microspheres under uniaxial compression,}
  J. Appl. Phys. \textbf{101}, 4908 (2007).

\bibitem{chen2006pressure}
Y.~Chen, A.~Best, H.-J. Butt, R.~Boehler, T.~Haschke, and W.~Wiechert,
  \enquote{Pressure distribution in a mechanical microcontact,} Appl. Phys.
  Lett. \textbf{88}, 234101 (2006).

\end{thebibliography}
\end{document}